\documentclass[12pt]{iopart}
\usepackage{iopams}  

\eqnobysec

\begin{document}

\newtheorem{proposition}{Proposition}


\renewcommand\vec[1]{\boldsymbol{#1}}
\newcommand\mymatrix[1]{\mathsf{#1}}

\newcommand\myset[1]{%
  \ifx #1\alpha \mathrm{A} \else 
  \ifx #1\beta  \mathrm{B} \else 
  \ifx #1\gamma \mathrm{C} \else 
  \ifx #1\delta \mathrm{D} \else 
  \ifx #1\kappa \mathrm{K} \else 
  \ifx #1\mu    \mathrm{M} \else 
  \ifx #1\xi    \mathrm{X} \else 
  \ifx #1\eta   \mathrm{Y} \else 
  \ifx #1\zeta  \mathrm{Z} \else 
  \boldsymbol{#1} \fi\fi\fi\fi\fi\fi\fi\fi\fi}

\newcommand\mysetsize[1]{|\myset{#1}|}

\newcommand\myT[1]{\mathbb{T}_{#1}}
\newcommand\myE[1]{\mathbb{E}_{#1}}
\newcommand\myu{u}
\newcommand\myw{w}

\newcommand\myShifted[3]{%
  \ifcase #1 \mathop{#2[#3]}%
  \or \myT{#3}#2%
  \or (\myT{#3}#2)%
  \else ?#1?#2[#3] \fi}
\newcommand\mymiwa[2]{(\mathbb{E}_{#1}{#2}) }

\newcommand\myOmega[1]{\if ?#1? \Omega\, \else \myShifted0{\Omega}{#1} \fi}
\newcommand\myPsi{\Phi}


\title{Soliton Fay identities. I. Dark soliton case.}
\author{V.E. Vekslerchik}
\address{
  Usikov Institute for Radiophysics and Electronics \\
  12, Proskura st., Kharkov, 61085, Ukraine 
}
\ead{vekslerchik@yahoo.com}
\submitto{\JPA}

\begin{abstract}
We derive a set of bilinear identities for the determinants of 
the matrices that have been used to construct 
the dark soliton solutions for various models.
To give examples of the application of the obtained identities we present 
soliton solutions for the equations describing multidimensional quadrilateral 
lattices, Darboux equations and multidimensional multicomponent systems of the 
nonlinear Schr\"odinger type.
\end{abstract}

\section{Introduction.}

It has been known since the 1970s that solutions of various integrable equations 
possess similar structure. Probably the brightest illustration of this fact is 
the so-called direct approach elaborated by Hirota, who i) noted that integrable 
equations can be reduced to a special, bilinear, form and ii) proposed the famous 
\textit{ansatz} to solve the latter. The universality of this \textit{ansatz} 
even gave rise to the search for other equations that can be solved in a 
similar way, which successfully resulted in the discovery of many new integrable 
models.
Later, similar approach, when one starts from the structure of solutions 
instead of, for example, constructing the inverse scattering transform, 
provided a wide range of classes of particular solutions built from 
determinants (Wronskians, Casoratians, determinants of Toeplitz or Hankel 
matrices, \textit{etc}) or Pfaffians. 
A large number of works is devoted to the (quasi)periodic solutions of 
integrable systems. 
Most of them are based on a modification of the inverse scattering method or 
the so-called algebro-geometric approach. The last method, which has been 
developed for almost all known integrable systems, exploits some rather 
sophisticated pieces of the theory of functions of complex variables. 
However, in many cases these solutions can be found with less effort, 
in a more straightforward way. One can find in \cite{Mumford} or 
\cite{V99,V00,PV10,V13a} examples of how to construct the `finite-genus' 
explicit solutions by means of the so-called Fay identities 
\cite{Mumford,Fay} for the $\theta$-functions associated with compact Riemann 
surfaces using calculations that seem elementary if compared with the 
`full-scale' algebro-geometric approach.

In all of the above cases we have different realizations of the direct 
approach that can be summarized as follows. 
One starts with the known structure of the 
solution (Hirota ansatz, combinations of Wronski/Casorati/Toeplitz/Hankel 
determinants or $\theta$-functions). Next, one has to determine some parameters 
and to demonstrate that this choice of parameters meets the equation in 
question. To do this one uses some identities related to the objects involved 
in the \textit{ansatz} (the identities from Hirota calculus, `classical' 
determinant identities or Fay identities for the $\theta$-functions).

The aim of this work is to simplify the direct approach for solitons. 
We study some properties of the matrices that have been used in 
\cite{V13a,V05,PV11,V11,V12a,V13b} to construct the dark soliton solutions 
for various models and derive some identities that are satisfied by these 
matrices and that can be viewed as soliton analogues of the Fay identities.

In the title of the paper (and throughout its text) we use the term `dark 
soliton' which is usually associated with localized objects existing 
against plane wave backgrounds or, from the mathematical viewpoint, 
solutions satisfying non-vanishing finite-density (\textit{i.e.} with 
constant or bounded modulus) boundary conditions. This usage is somewhat 
dubious because we, first, do not discuss below the boundary value problems. 
Second, in examples in section \ref{sec-app} we present solutions with 
unbounded background (though they can be made bounded by admitting complex 
parameters, the question that, again, we do not discuss in this paper). 
Finally, our approach may lead to the so-called anti-dark solitons. 
Nevertheless, we use this term 
i) to indicate that the identities which are particular cases of the 
Fay identities presented below have been used to obtain dark soliton 
solutions for various models and ii) to distinguish the case of this paper 
from the bright soliton case (with zero boundary conditions) where soliton 
matrices/determinants and, hence, Fay identities have similar but different 
structures and which will be studied in the second part of this work.

We start with introducing in section \ref{sec-mat} a class of 
matrices than can be called `dark soliton matrices'. 
By simple algebra we deduce the set of \textit{bilinear} identities that are 
satisfied by the determinants of these matrices (section \ref{sec-fay}), 
present their alternative formulations (section \ref{sec-miwa}) and their 
differential versions (section \ref{sec-diff}). 
We discuss a few applications of the obtained identities in section 
\ref{sec-app}, where we use them to construct solutions for the equations 
describing multidimensional lattices, Darboux equations and multicomponent 
equations of the nonlinear Schr\"odinger (NLS) type.

\section{Dark soliton matrices. \label{sec-mat}}

\subsection{Notation}

The central objects of our study are the determinants 
\begin{equation}
  \Omega(\mymatrix{A}) = 
  \det\left| \mymatrix{1} + \mymatrix{A} \right|
\label{def-omega}
\end{equation}
and the shifts that are defined as
\begin{equation}
  \myShifted1{\Omega(\mymatrix{A})}{\zeta} = 
  \det\left| \mymatrix{1} + \mymatrix{A}\mymatrix{H}_{\zeta} \right|
\label{def-T} 
\end{equation} 
where $\mymatrix{H}_{\zeta}$  is the constant diagonal matrix defined by, 
\begin{equation}
  \mymatrix{H}_{\zeta} = 
  \left( \zeta - \mymatrix{L} \right)
  \left( \zeta - \mymatrix{R} \right)^{-1}
\label{def-H} 
\end{equation}
with constant diagonal matrices $\mymatrix{L}$, $\mymatrix{R}$ and
unit matrices being omitted: 
$\zeta - \mymatrix{L} $ stands for $\zeta\mymatrix{1} - \mymatrix{L}$ etc.

To simplify the following formulae we use the `set' notation, 
$ \myset{\xi} = \{ \xi_{1}, ..., \xi_{n} \}$, 
with $|...|$ standing for the number of elements of the set, 
$ \mysetsize{\xi} = n $, 
and basic set operations denoted as 
$\myset\xi \myset\eta = \myset\xi \cup \myset\eta$, 
$\zeta\myset{\xi} = \myset{\xi}\zeta = \{\zeta\} \cup \myset{\xi}$  
and 
$  \myset{\xi}/\xi_{k} 
  = 
  \{ \xi_{1}, ... , \widehat{\xi_{k}}, ... \xi_{n} \}
$.
We use the set indices to indicate the superposition of the shifts, 
\begin{equation}
  \myShifted1{}{\myset{\xi}} 
  = 
  \prod_{\xi \in \myset{\xi} } \myShifted1{}{\xi} 
\end{equation}
or 
\begin{equation}
  \myShifted1{\Omega(\mymatrix{A})}{\myset{\xi}} 
  = 
  \det\left| \mymatrix{1} + \mymatrix{A}\mymatrix{H}_{\myset{\xi}} \right|
\end{equation}
where 
\begin{equation}
  \mymatrix{H}_{\myset{\xi}} = 
  \prod_{\xi \in \myset{\xi} } \mymatrix{H}_{\xi}. 
\end{equation}
Sometimes, when the matrix $\mymatrix{A}$ is fixed, we do not write the 
$\mymatrix{A}$-dependence of $\myOmega{}$ explicitly and use an alternative 
notation for the shifted determinants, 
\begin{equation}
  \myOmega{\myset{\xi}} = 
  \myShifted1{\myOmega{}}{\myset{\xi}} = 
  \det\left| \mymatrix{1} + \mymatrix{A}\mymatrix{H}_{\myset{\xi}} \right|. 
\label{not-shft}
\end{equation}

\subsection{Rank one condition.}

In the following sections we derive the identities for the determinants of the 
so-called `almost-intertwining' matrices \cite{KG01} that satisfy the 
`rank one condition' \cite{GK02,GK06,NAH09}, which is a particular case of 
the Sylvester equation (see \cite{DM10a,DM10b}). We write this condition as 
\begin{equation}
  \mymatrix{L} \mymatrix{A} - \mymatrix{A} \mymatrix{R} 
  = 
  | \,\ell\, \rangle \langle a |.
\label{cond-rank-one}
\end{equation}
Here, 
$| \,\ell\, \rangle$ is a constant $N$-component column, 
$| \,\ell\, \rangle = \left( \ell_{1}, ... , \ell_{N} \right)^{T}$, 
$\langle a |$ is an $N$-component row, 
$\langle a | = \left( a_{1}, ... , a_{N} \right)$. 
From the viewpoint of applications, namely vectors $\langle a |$ determine 
the dependence of the matrices $\mymatrix{A}$ on the coordinates describing 
the model we are dealing with.

One of the consequences of \eref{cond-rank-one} is the following 
\begin{proposition}
\label{prop-PN}
The shifted determinants \eref{def-omega} can be presented as 
\begin{eqnarray}
  \myT\zeta \Omega(\mymatrix{A}) 
  & = & 
  \Omega(\mymatrix{A}) 
  \left[
  1 + 
  \langle a | (\mymatrix{R} - \zeta)^{-1} \mymatrix{G}(\mymatrix{A}) | \,\ell\, \rangle
  \right], 
\label{eq-omega-p}
\\[2mm]
  \myT\zeta^{-1} \Omega(\mymatrix{A}) 
  & = & 
  \Omega(\mymatrix{A}) 
  \left[
  1 - 
  \langle a | \mymatrix{G}(\mymatrix{A}) (\mymatrix{L} - \zeta)^{-1} | \,\ell\, \rangle
  \right]
\label{eq-omega-n}
\end{eqnarray}
where 
\begin{equation}
  \mymatrix{G}(\mymatrix{A}) = ( \mymatrix{1} + \mymatrix{A} )^{-1}.  
\end{equation}
\end{proposition}
The proof of these statements is presented in \ref{proof-PN}. 

The following result plays a crucial role in the derivation of the soliton 
Fay identities.
\begin{proposition} 
\label{prop-AB} 
If matrices $\mymatrix{X}$ and $\mymatrix{Y}$ are given by 
\begin{equation}
  \mymatrix{X} = \mymatrix{A} \mymatrix{H}_{\myset\xi}, 
  \qquad
  \mymatrix{Y} = \mymatrix{A} \mymatrix{H}_{\myset\eta} 
\end{equation}
where $\myset\xi$ and $\myset\eta$ are two arbitrary sets, then  
\begin{equation}
  \left(\myT\zeta \Omega(\mymatrix{X}) \right)
  \left(\myT\zeta^{-1} \Omega(\mymatrix{Y}) \right)
  = 
  \Omega(\mymatrix{X}) \Omega(\mymatrix{Y}) 
  \left[
  1 
  + 
  \langle y | \mymatrix{G}(\mymatrix{Y}) 
  \mathcal{D}_{\myset{\xi}\myset{\eta}}(\zeta) 
  \mymatrix{G}(\mymatrix{X}) 
  | \,\ell\, \rangle
  \right]
\label{dets-XY}
\end{equation}
with $\langle y |  = \langle a |  \mymatrix{H}_{\myset\eta}$ and 
\begin{equation}
  \mathcal{D}_{\myset\xi \myset\eta}(\zeta) =
  \mymatrix{H}_{\myset\xi} \mymatrix{H}_{\myset\eta}^{-1} 
  \left(\mymatrix{R} - \zeta \right)^{-1} 
  - 
  \left(\mymatrix{L} - \zeta \right)^{-1}
\label{Delta-XY}
\end{equation}
\end{proposition}
(see \ref{proof-AB} for a proof).

\subsection{Derivation of the main identity.}

The significance of proposition \ref{prop-AB} is that it `extracts' the dependence on 
$\zeta$ outside the determinants and demonstrates that the left-hand side of 
\eref{dets-XY} is a meromorphic function of $\zeta$.
Now we use the elementary formula for the quotient of the polynomials,
\begin{equation}
  \frac{ Q(t) }{ P(t) }
  = 
  \sum_{i=1}^{n} \frac{ c_{i} }{ t - \eta_{i} } 
\end{equation}
where $P(t) = \prod_{i=1}^{n} (t - \eta_{i})$, $Q(t)$ is a polynomial whose 
degree is less than $n$ and $c_{i} = Q(\eta_{i}) / P'(\eta_{i})$ 
which we rewrite as 
\begin{equation}
  \frac{ \Delta(t, \myset\xi) }{ \Delta(t, \myset\eta) } 
  = 
  \sum_{\eta \in \myset\eta}
  \frac{ \Gamma_{\eta}(\myset\xi, \myset\eta) }{ t-\eta }, 
  \qquad
  \Gamma_{\eta}(\myset\xi, \myset\eta) 
  = 
  \frac{ \Delta(\eta, \myset\xi) }{ \Delta( \eta, \myset\eta/\eta) }
\label{interpolation}
\end{equation}
with
\begin{equation}
  \Delta(t,\myset{\eta}) = 
  \prod_{\eta \in \myset{\eta} } 
  (t - \eta), 
\qquad
  \Delta(\myset{\xi},\myset{\eta}) = 
  \prod_{\xi  \in \myset{\xi} } 
  \prod_{\eta \in \myset{\eta} } 
  (\xi - \eta). 
\end{equation}
It follows from \eref{interpolation} and \eref{Delta-XY} that, in the case of 
$\mysetsize\xi<\mysetsize\eta$,
\begin{equation}
\fl \qquad
  \sum_{\zeta \in \myset\eta}
  \Gamma_{\zeta}(\myset\xi, \myset\eta) 
  \mathcal{D}_{\myset\xi \myset\eta}(\zeta) 
  = 
  \mymatrix{H}_{\myset\xi} 
  \mymatrix{H}_{\myset\eta}^{-1} 
  \Delta(\mymatrix{R}, \myset\xi ) 
  \Delta^{-1}(\mymatrix{R}, \myset\eta ) 
  - 
  \Delta(\mymatrix{L}, \myset\xi ) 
  \Delta^{-1}(\mymatrix{L}, \myset\eta )
\end{equation}
which leads, together with 
$\mymatrix{H}_{\myset\xi}\Delta(\mymatrix{R},\myset\xi )=\Delta(\mymatrix{L},\myset\xi)$, 
to
\begin{equation}
  \sum_{\zeta \in \myset\eta}
  \Gamma_{\zeta}(\myset\xi, \myset\eta) 
  \mathcal{D}_{\myset\xi \myset\eta}(\zeta) 
  = 0. 
\end{equation}
Another consequence of \eref{interpolation} 
is that in the case of $\mysetsize\xi<\mysetsize\eta$
\begin{equation}
  \sum_{\zeta \in \myset\eta}
  \Gamma_{\zeta}(\myset\xi, \myset\eta) 
  = 
  \delta_{ \mysetsize{\xi}+1,\mysetsize{\eta} } 
\end{equation}
which, together with \eref{dets-XY}, implies 
\begin{equation}
  \sum_{\zeta \in \myset\eta}
  \Gamma_{\zeta}(\myset\xi, \myset\eta) 
  \left(\myT\zeta \Omega(\mymatrix{X}) \right)
  \left(\myT\zeta^{-1} \Omega(\mymatrix{Y}) \right)
  = 
  \delta_{ \mysetsize{\xi}+1,\mysetsize{\eta} } 
  \Omega(\mymatrix{X}) \Omega(\mymatrix{Y}). 
\end{equation}
Noting that 
\begin{equation}
  \myT\zeta \Omega(\mymatrix{X}) = 
  \myShifted1{\Omega(\mymatrix{A})}{\zeta\myset\xi}
\end{equation} 
and 
\begin{equation}
  \myT\zeta^{-1} \Omega(\mymatrix{Y}) =
  \myShifted1{\Omega(\mymatrix{A})}{\myset\eta/\zeta}  
  \quad \mbox{for} \quad \zeta \in \myset\eta
\end{equation}
one can rewrite the last identity as 
\begin{equation}
  \sum_{\eta \in \myset\eta}
  \Gamma_{\eta}(\myset\xi, \myset\eta) 
  \left(\myT{\eta\myset\xi}   \Omega(\mymatrix{A}) \right)
  \left(\myT{\myset\eta/\eta} \Omega(\mymatrix{A}) \right)
  = 
  \delta_{ \mysetsize{\xi}+1,\mysetsize{\eta} } 
  \left(\myT{\myset\xi}  \Omega(\mymatrix{A}) \right)
  \left(\myT{\myset\eta} \Omega(\mymatrix{A}) \right). 
\label{eq-main}
\end{equation}

\section{Fay identities. \label{sec-fay}}

\subsection{The main identity.}

Hereafter, we fix the `base' matrix $\mymatrix{A}$, do not write it explicitly 
and use notation \eref{not-shft}. Statement \eref{eq-main}, which is the main 
identity of this paper, can be formulated as 
\begin{proposition}
\label{prop-main}
For any sets $\myset\xi$ and $\myset\eta$, 
$\mysetsize\xi<\mysetsize\eta$, 
the shifted determinants $\myOmega{}$, 
\begin{equation}
  \myOmega{\myset{\xi}} = 
  \det\left| \mymatrix{1} + \mymatrix{A}\mymatrix{H}_{\myset{\xi}} \right|, 
\end{equation}
satisfy the following bilinear identity: 
\begin{equation}
  \delta_{ \mysetsize{\xi}+1,\mysetsize{\eta} } 
  \myOmega{\myset\xi} \myOmega{\myset\eta}  
  = 
  \sum_{\eta  \in \myset{\eta} } 
  \Gamma_{\eta}(\myset\xi,\myset\eta) \, 
  \myOmega{\myset{\xi}\eta} \myOmega{\myset{\eta}/\eta} 
\label{Fay-main}
\end{equation}
where 
\begin{equation}
  \Gamma_{\eta}(\myset\xi,\myset\eta) 
  = 
  \frac{ \Delta(\eta, \myset\xi) }{ \Delta(\eta, \myset{\eta}/\eta) }. 
\label{def-Gamma}
\end{equation}
\end{proposition}

It is easy to see that equation \eref{Fay-main} is a straightforward 
generalization of the famous Hirota bilinear difference equation \cite{H81} 
(when $\mysetsize{\xi}+1 < \mysetsize{\eta}$) or the Miwa equation \cite{M82}  
(when $\mysetsize{\xi}+1 = \mysetsize{\eta}$). 
Thus, proposition \ref{prop-main} states that we have obtained the dark 
soliton solutions for the Hirota and Miwa equations.

\subsection{Subset identities.}

To make visible the `combinatorial' aspects of the Fay identities, it is 
instructive to introduce the function
\begin{equation}
  \myPsi(\myset{\xi},\myset{\eta}) = 
  \frac{ \myOmega{\myset{\xi}} \myOmega{\myset{\eta}} } 
       { \Delta(\myset{\xi},\myset{\eta}) }.
\end{equation}
In terms of $\myPsi$, which has the property 
\begin{equation}
  \myPsi(\myset{\xi},\myset{\eta}) = 
  (-)^{ \mysetsize{\xi}\mysetsize{\eta} } 
  \myPsi(\myset{\eta},\myset{\xi}), 
\end{equation}
the main identity \eref{Fay-main} looks like 
\begin{equation}
  (-)^{ \mysetsize{\xi} } 
  \delta_{ \mysetsize{\xi}+1,\mysetsize{\eta} } 
  \myPsi(\myset{\xi},\myset{\eta}) 
  = 
  \sum_{\eta  \in \myset{\eta} } 
  \myPsi(\myset{\xi}\eta,\myset{\eta}/\eta)
\qquad
  (\mysetsize{\xi} < \mysetsize{\eta} ). 
\end{equation}

Choosing some set $\myset\zeta$ as `universal' one and taking complementary 
$\myset\xi$ and $\myset\eta$, $\myset\xi \cup \myset\eta = \myset\zeta$, one 
can derive from the above equation that, if $n$ meets 
$N+1 \leqslant 2n$, $n \leqslant N$ 
where $N = \mysetsize\zeta$, then  
\begin{equation}
  (-)^{N-n} \delta_{2n,N+1} 
  \sum_{\stackrel{\scriptstyle\myset\eta \subset \myset\zeta}{\mysetsize\eta=n}}
  \chi(\myset\eta) 
  \myPsi(\myset\zeta/\myset\eta, \myset\eta)
  = 
  \sum_{\stackrel{\scriptstyle\myset\xi \subset \myset\zeta}{\mysetsize\xi=n-1}}
  \chi_{1}(\myset\zeta, \myset\xi) 
  \myPsi(\myset\zeta/\myset\xi, \myset\xi)
\label{chi-a}
\end{equation}
with 
\begin{equation}
  \chi_{1}(\myset\zeta,\myset\xi) 
  = 
  \sum_{ \zeta \in \myset\zeta/\myset\xi }
  \chi(\zeta\myset\xi). 
\end{equation}
Analyzing \eref{chi-a} in the case when $\chi$ is the characteristic function, 
$\chi(\myset\xi)=\mysetsize\xi$, one can deduce 
\begin{equation}
  \sum_{\stackrel{\scriptstyle\myset\xi \subset \myset\zeta}{\mysetsize\xi=n}}
  \myPsi(\myset\zeta/\myset\xi, \myset\xi)
  = 0 
  \qquad
  (
  1 \leqslant n \leqslant N-1,
  \quad
  N \geqslant 2
  ).
\label{chi-b}
\end{equation} 
A more general result can be obtained for any additive $\chi$, 
$\chi(\myset\xi\myset\eta)=\chi(\myset\xi)+\chi(\myset\eta)$.
In this case 
\begin{equation}
  \chi_{1}(\myset\zeta, \myset\xi) 
  = 
  \chi(\myset\zeta) 
  + 
  \left( \mysetsize\zeta - \mysetsize\xi - 1 \right)
  \chi(\myset\xi) 
\end{equation}
and \eref{chi-a} together with \eref{chi-b} imply that function 
\begin{equation}
  F_{n} = 
  \sum_{\stackrel{\scriptstyle\myset\xi \subset \myset\zeta}{\mysetsize\xi=n}}
  \chi(\myset\xi) 
  \myPsi(\myset\zeta/\myset\xi, \myset\xi) 
\end{equation}
satisfies equation 
\begin{equation}
  (-)^{N-n} \delta_{2n,N+1} F_{n} = 
  (N-n) F_{n-1} 
  \qquad 
  \left(N/2 < n \leqslant N \right). 
\end{equation}
From this equations and 
\begin{equation}
  F_{N-n} 
  = 
  (-)^{n(N-1)+1} F_{n}, 
\end{equation}
which follows from the definition of $F_{n}$, one can easily obtain that 
$F_{n}=0$ for $2 \leqslant n \leqslant N-2$, or 
\begin{equation}
  \sum_{\stackrel{\scriptstyle\myset\xi \subset \myset\zeta}{\mysetsize\xi=n}}
  \chi(\myset\xi) 
  \myPsi(\myset\zeta/\myset\xi, \myset\xi) 
  = 0 
  \qquad
  (
  2 \leqslant n \leqslant N-2, 
  \quad 
  N \geqslant 4
  ).
\end{equation}
Returning to the determinants $\myOmega{\myset\xi}$, 
one arrives at 
\begin{proposition} 
For any set $\myset\zeta$ and arbitrary function $\chi(\xi)$, 
the shifted determinants $\myOmega{}$  
satisfy the following bilinear identities: 
\begin{eqnarray}
  \sum_{\stackrel{\scriptstyle\myset\xi \subset \myset\zeta}{\mysetsize\xi=n}}
  \gamma_{1}(\myset\xi,\myset\zeta) \, 
  \myOmega{\myset{\xi}} \myOmega{\myset\zeta/\myset\xi} 
  = 0 
  &\qquad&
  (1 \leqslant n \leqslant \mysetsize\zeta-1, \mysetsize\zeta \geqslant 2) 
\\
  \sum_{\stackrel{\scriptstyle\myset\xi \subset \myset\zeta}{\mysetsize\xi=n}}
  \gamma_{\chi}(\myset\xi,\myset\zeta) \, 
  \myOmega{\myset{\xi}} \myOmega{\myset\zeta/\myset\xi} 
  = 0 
  &\qquad&
  (2 \leqslant n \leqslant \mysetsize\zeta-2, \mysetsize\zeta \geqslant 4) 
\end{eqnarray}
where 
\begin{equation}
  \gamma_{1}(\myset\xi,\myset\zeta) 
  = 
  \frac{ 1 }
       { \Delta(\myset\xi, \myset\zeta/\myset\xi) }, 
\qquad
  \gamma_{\chi}(\myset\xi,\myset\zeta) 
  = 
  \frac{ \sum_{\xi \in \myset\xi} \, \chi(\xi) }
       { \Delta(\myset\xi, \myset\zeta/\myset\xi) }. 
\end{equation}
\end{proposition}

\section{Miwa shifts. \label{sec-miwa}}

One of the methods used to tackle integrable models is the so-called functional 
representation when an infinite set of integrable equations (a hierarchy) is 
presented as a functional equation or a finite set of functional equations 
which  are usually written in terms of the Miwa shifts. 
These shifts are defined by 
\begin{equation}
  \myE\zeta 
  \left. f( t_{k} )\right|_{k=1}^{\infty} 
  = 
  \left. f( t_{k} + \epsilon \zeta^{k}/k )\right|_{k=1}^{\infty} 
\end{equation}
where $\epsilon$ is a constant (usually, $1$ of $i$). 

From the definition of the matrices $\mymatrix{H}_{\alpha}$ \eref{def-H} 
it is clear that the shifts $\myT\alpha$ can play the role of the Miwa shifts, or 
give rise to the continuous flows, only near the point $\alpha=\infty$ 
(otherwise one cannot meet the condition $\lim_{\zeta \to 0}\myE\zeta=1$), 
which restricts universality of the presented Fay identities. 
More flexible approach is to model the Miwa shifts as 
\begin{equation}
  \mathbb{E}_{\alpha} 
  = 
  \myT\alpha \myT\kappa^{-1} 
\label{def-miwa}
\end{equation}
where $\kappa$ is some fixed (constant) parameter, that leads to the typical 
presentation of $\mathbb{E}_{\alpha}$ as a power series 
$
  \mathbb{E}_{\alpha} 
  = 
  1 +
  \sum_{n=1}^{\infty} (\alpha - \kappa)^{n} 
  \mathcal{D}_{n}  
$
where $\mathcal{D}_{n}$ are some differential operators. 

In this section we rewrite the main identity \eref{Fay-main} and its 
consequences in terms of the operators \eref{def-miwa} or, more generally, 
\begin{equation}
  \myE{\myset\alpha} 
  = 
  \prod_{\alpha \in \myset\alpha} \mathbb{E}_{\alpha} 
  = 
  \myT{\myset\alpha} 
  \myT{\kappa}^{-\mysetsize\alpha}. 
\label{def-Miwa}
\end{equation}
Considering various ways of constructing sets $\myset\xi$ and $\myset\eta$ 
of 
$\myset\kappa_{n} = \{ \overbrace{\kappa ... \kappa}^{n\,\mathrm{times}} \}$, 
$\myset\alpha$ and $\myset\beta$, with 
$\kappa \notin \myset\alpha,\myset\beta$ 
one can derive various $\myE{}$-identities.
For example, taking $\myset\xi=\myset\beta$, $\myset\eta=\kappa\myset\alpha$ 
with $\mysetsize\alpha=\mysetsize\beta + n$, $n=1,2,3$ 
one can rewrite \eref{Fay-main}  as 
\begin{eqnarray}
  \mymiwa{\myset\alpha}{\Omega} 
  \mymiwa{\myset\beta}{\Omega} 
  & = & 
  \sum_{ \alpha \in \myset\alpha } 
  \gamma_{\alpha}( \myset\alpha, \myset\beta) \; 
  \mymiwa{\myset\alpha/\alpha}{\Omega} 
  \mymiwa{\alpha\myset\beta}{\Omega} 
\qquad (\mysetsize\alpha=\mysetsize\beta + 1)
\label{Miwa-a}
\\
  \mymiwa{\myset\alpha}{\hat\Omega} 
  \mymiwa{\myset\beta}{\Omega} 
  & = & 
  \sum_{ \alpha \in \myset\alpha } 
  \gamma_{\alpha}( \myset\alpha, \myset\beta) \; 
  \mymiwa{\myset\alpha/\alpha}{\hat\Omega} 
  \mymiwa{\alpha\myset\beta}{\Omega} 
\qquad (\mysetsize\alpha=\mysetsize\beta + 2)
\\
  \mymiwa{\myset\alpha}{\Omega} 
  \mymiwa{\myset\beta}{\check\Omega} 
  & = & 
  \sum_{ \alpha \in \myset\alpha } 
  \gamma_{\alpha}( \myset\alpha, \myset\beta) \; 
  \mymiwa{\myset\alpha/\alpha}{\Omega} 
  \mymiwa{\alpha\myset\beta}{\check\Omega} 
\qquad (\mysetsize\alpha=\mysetsize\beta + 2)
\\
  \mymiwa{\myset\alpha}{\hat\Omega} 
  \mymiwa{\myset\beta}{\check\Omega} 
  & = & 
  \sum_{ \alpha \in \myset\alpha } 
  \gamma_{\alpha}( \myset\alpha, \myset\beta) \; 
  \mymiwa{\myset\alpha/\alpha}{\hat\Omega} 
  \mymiwa{\alpha\myset\beta}{\check\Omega} 
\qquad (\mysetsize\alpha=\mysetsize\beta + 3)
\end{eqnarray}
where 
\begin{equation}
  \hat\Omega = \myT\kappa \Omega, 
\qquad
  \check\Omega = \myT\kappa^{-1}\Omega
\label{def-hat}
\end{equation}
and
\begin{equation}
  \gamma_{\alpha}( \myset\alpha, \myset\beta) 
  =
  \frac{ \Delta(\alpha, \myset\beta) }
       { \Delta(\alpha, \myset\alpha/\alpha) } \, 
  \frac{ \Delta(\kappa, \myset\alpha/\alpha) }
       { \Delta(\kappa, \myset{\beta}) }.  
\end{equation}
From the above formulae one can obtain a set of `homogeneous' equations by 
replacing $\myset\beta$ with $\myset\beta \myset\kappa_{m}$: 
\begin{eqnarray}
  0 & = & 
  \sum_{ \alpha \in \myset\alpha } 
  \gamma_{\alpha}( \myset\alpha, \myset\beta) 
  (\alpha-\kappa)^{\mysetsize\alpha - \mysetsize\beta - 1}  
  \mymiwa{\myset\alpha/\alpha}{\Omega} 
  \mymiwa{\alpha\myset\beta}{\Omega} 
\quad (\mysetsize\alpha > \mysetsize\beta + 1), 
\\ 
  0 & = & 
  \sum_{ \alpha \in \myset\alpha } 
  \gamma_{\alpha}( \myset\alpha, \myset\beta) 
  (\alpha-\kappa)^{\mysetsize\alpha - \mysetsize\beta - 2}  
  \mymiwa{\myset\alpha/\alpha}{\hat\Omega} 
  \mymiwa{\alpha\myset\beta}{\Omega} 
\quad (\mysetsize\alpha > \mysetsize\beta + 2), 
\\ 
  0 & = & 
  \sum_{ \alpha \in \myset\alpha } 
  \gamma_{\alpha}( \myset\alpha, \myset\beta) 
  (\alpha-\kappa)^{\mysetsize\alpha - \mysetsize\beta - 2}  
  \mymiwa{\myset\alpha/\alpha}{\Omega} 
  \mymiwa{\alpha\myset\beta}{\check\Omega} 
\quad (\mysetsize\alpha > \mysetsize\beta + 2),
\\
  0 & = & 
  \sum_{ \alpha \in \myset\alpha } 
  \gamma_{\alpha}( \myset\alpha, \myset\beta) 
  (\alpha-\kappa)^{\mysetsize\alpha - \mysetsize\beta - 3}  
  \mymiwa{\myset\alpha/\alpha}{\hat\Omega} 
  \mymiwa{\alpha\myset\beta}{\check\Omega} 
\quad (\mysetsize\alpha > \mysetsize\beta + 3).  
\end{eqnarray}
In a similar way, \eref{Fay-main} with $\myset\xi=\myset\beta$, 
$\myset\eta=\kappa\myset\alpha$ and $\mysetsize\alpha=\mysetsize\beta$ leads, 
after replacing $\myset\beta \to \myset\beta \myset\kappa_{n}$
and applying the shift $\myT\kappa^{-1}$, to 
\begin{eqnarray}
&&
  \delta_{\mysetsize\alpha, \mysetsize\beta}
  \mymiwa{\myset\alpha}{\Omega} 
  \mymiwa{\myset\beta}{\hat\Omega} 
  - 
  \frac{ \Delta(\kappa, \myset\alpha) }
       { \Delta(\kappa, \myset\beta) } \; 
  \mymiwa{\myset\alpha}{\hat\Omega} 
  \mymiwa{\myset\beta}{\Omega} 
\nonumber\\&&\qquad
  = 
  \sum_{ \alpha \in \myset\alpha } 
  \gamma_{\alpha}( \myset\alpha, \myset\beta) 
  (\alpha-\kappa)^{\mysetsize\alpha - \mysetsize\beta}  
  \mymiwa{\myset\alpha/\alpha}{\Omega} 
  \mymiwa{\alpha\myset\beta}{\hat\Omega} 
\\&&
  \delta_{\mysetsize\alpha, \mysetsize\beta}
  \mymiwa{\myset\alpha}{\check\Omega} 
  \mymiwa{\myset\beta}{\Omega} 
  - 
  \frac{ \Delta(\kappa, \myset\alpha) }
       { \Delta(\kappa, \myset\beta) } \; 
  \mymiwa{\myset\alpha}{\Omega} 
  \mymiwa{\myset\beta}{\check\Omega} 
\nonumber\\&&\qquad
  =
  \sum_{ \alpha \in \myset\alpha } 
  \gamma_{\alpha}( \myset\alpha, \myset\beta) 
  (\alpha-\kappa)^{\mysetsize\alpha - \mysetsize\beta}  
  \mymiwa{\myset\alpha/\alpha}{\check\Omega} 
  \mymiwa{\alpha\myset\beta}{\Omega} 
\end{eqnarray} 
for $\mysetsize\alpha \geqslant \mysetsize\beta$.

Looking at the above equations, one can see the appearance of the triplet 
$\check\Omega$, $\Omega$, $\hat\Omega$ instead of $\Omega$. 
This situation is typical for the `complex' NLS-like models. 
Indeed, these models are formulated in terms of two functions $q$ and $r$, 
which are related in physical applications by $r = \pm\bar{q}$ 
(bar stands for the complex conjugation). To bilinearize NLS-like equations 
one usually has to employ three tau-functions $\sigma$, $\tau$ and $\rho$: 
$q=\sigma/\tau$, $r = \rho/\tau$ with involution $\rho=\pm\bar\sigma$ and 
$\tau=\bar\tau$. 
It was shown in the previous works by the author that in the dark soliton case 
the universal way to construct the chain (finite or infinite) of the tau 
functions $ ... \to \rho \to \tau \to \sigma \to ...$ 
is to use the $\myT\mu$ shifts, 
$\tau \propto \Omega$, 
$\rho \propto \myT\mu^{-1}\Omega$,
$\sigma \propto \myT\mu\Omega$, \textit{etc}, with some distinguished $\mu$.  
In particular, the case of this section, $\mu=\kappa$ where $\kappa$ is the 
`base' of the Miwa shifts, corresponds, as is shown in \cite{V12a}, to the 
the positive (classical) part of the AKNS hierarchy. 

Construction \eref{def-Miwa} can be extended to introduce other sets of Miwa 
shifts by changing the `base' shift parameter $\kappa$. In the previous works 
the author used two sets, 
\begin{equation}
  \myE{\alpha} 
  = 
  \myT{\alpha} 
  \myT{\kappa}^{-1}, 
\qquad
  \overline\myE{\beta} 
  = 
  \myT{\beta} 
  \myT{\bar\kappa}^{-1}, 
\qquad
  \kappa \ne \bar\kappa
\end{equation}
to deal with extended hierarchies consisted of positive, $\myE{}$, 
and negative, $\overline\myE{}$, subhierarchies.

More generally, one can introduce several Miwa shifts and, hence, several 
systems of continuous flows, 
$\myE{\alpha}^{(i)} = \myT{\alpha} \myT{\kappa_{i}}^{-1}$, 
to deal with multidimensional equations and hierarchies.

We do not present here the corresponding identities because it seems difficult 
to select from possible variants some reasonable set of `ready-to-use' 
formulae.

\section{Differential Fay identities. \label{sec-diff}}

The differential Fay identities can be derived using the standard limit 
procedure. 
First we define the differential operator $\partial_{\lambda}$ by 
\begin{equation}
  \partial_{\lambda} = 
  \lim_{\xi\to\lambda} 
  \frac{ 1 }{ \xi - \lambda } 
  \left( \myT{\xi} \myT{\lambda}^{-1} - 1 \right). 
\label{partial-def}
\end{equation}
Then, considering, as in the case of the Miwa shifts, various ways of 
constructing sets $\myset\xi$ and $\myset\eta$ 
of $\lambda$, $\xi$, $\myset\alpha$ and $\myset\beta$, with 
$\lambda,\xi \notin \myset\alpha,\myset\beta$ and taking the 
$\xi\to\lambda$ limit one can obtain from the main identity \eref{Fay-main} 
various bilinear equations involving $\partial_{\lambda}$ or its bilinear 
analogue, 
\begin{equation}
  D_{\lambda} \, u \cdot v 
  = 
  \left( \partial_{\lambda} u \right) v 
  - 
  u \left( \partial_{\lambda} v \right). 
\end{equation}
We present here only the general one, that can be written 
for the case $\mysetsize{\alpha} + 1 \ge \mysetsize{\beta}$ 
as 
\begin{eqnarray}
\fl
  \frac{ \Delta( \lambda,  \myset\beta ) }
       { \Delta( \lambda, \myset\alpha ) }
  \biggl[ 
    D_{\lambda} 
    + \Lambda_{\lambda}(\myset\alpha) 
    - \Lambda_{\lambda}(\myset\beta) 
  \biggr] 
  \myOmega{\myset\alpha} \cdot \myOmega{\myset\beta} 
\nonumber
\\ 
  =  
  \sum_{ \alpha \in \myset\alpha } 
  \frac{ 1 }{ (\lambda-\alpha)^{2} } 
  \frac{ \Delta( \alpha, \myset\beta ) }
       { \Delta( \alpha, \myset\alpha/\alpha ) }
  \myOmega{ \lambda\myset\alpha/\alpha } 
  \myOmega{ \bar{\lambda}\alpha\myset\beta } 
  - 
  \delta_{\mysetsize\alpha + 1, \mysetsize\beta}
  \myOmega{ \lambda\myset\alpha } 
  \myOmega{ \bar{\lambda}\myset\beta } 
\label{Fay-diff}
\end{eqnarray}
where letters with overbar indicate inverse shifts,
\begin{equation}
  \myOmega{ \bar{\lambda}\myset\xi } 
  = 
  \myT\lambda^{-1} \myOmega{ \myset\xi } 
\end{equation}
and 
\begin{equation}
  \Lambda_{\lambda}(\myset\alpha) 
  = 
  \sum_{ \alpha \in \myset\alpha } 
  \frac{ 1 }{ \lambda - \alpha }
\end{equation}
together with the Toda-like identity of the second order,  
\begin{equation}
  D_{\lambda\mu} \, 
  \myOmega{} \cdot \myOmega{} 
  =  
  \frac{ 2 }{ (\lambda-\mu)^{2} } 
  \myOmega{ \lambda\bar\mu}\myOmega{ \mu\bar\lambda} 
  + 
  c \, \myOmega{}^{2}  
\label{eq-Toda}
\end{equation}
where $c$ is a constant that is usually determined from the boundary 
conditions.

From the practical viewpoint, the differential identities \eref{Fay-diff} are 
to be combined with the `algebraic' ones, \eref{Fay-main} or formulae from 
section \ref{sec-miwa}, to present the right-hand side of \eref{Fay-diff} 
as a `physical' combination of the tau functions $\Omega$.
One can find a few examples of application of differential Fay identities in 
the next section.

\section{Applications. \label{sec-app}}

In this section we discuss applications of the Fay identities derived in this 
paper.
We would like to note that particular cases of the formulae of the 
previous sections have already been used by the author in previous works 
\cite{V13a,V05,PV11,V11,V12a,V13b}, that may be considered as the 
`most practical' examples.

Here we focus on the models that can be viewed as extensions of the `classical' 
integrable models. The main feature of all equations presented below is that 
they are \textit{multidimensional}, contrary to most of the `classical', 
or integrable by means of the standard inverse scattering transform, models 
that are $2$- or  $(1+1)$-dimensional.

\subsection{Multidimensional lattices. \label{sec-mql}}

Consider the two-parametric set of functions $f_{\eta\zeta}$ 
\begin{equation}
  f_{\eta\zeta} = 
  \varphi_{\eta\zeta}  
  \frac{ \myOmega{\eta\bar\zeta} }{ \myOmega{} } 
\label{def-ql-f}
\end{equation}
where `background' functions $\varphi_{\eta\zeta}$ are defined by 
\begin{equation}
  \myShifted1{\varphi_{\eta\zeta}}{\myset\xi} 
  = 
  \frac{ \Delta(\eta, \myset{\xi}) }{ \Delta(\zeta, \myset{\xi}) } \; 
  \varphi_{\eta\zeta},
  \qquad  
  \varphi_{\eta\eta}=1. 
\end{equation}
It follows from the main identity \eref{Fay-main} with 
$\myset\xi \to \emptyset$ and $\myset\eta \to \eta\zeta\myset\xi$ 
after the shift $\myT\zeta^{-1}$ that 
\begin{equation}
  \left( \myT{\myset\xi} - 1 \right) f_{\eta\zeta}
  = 
  \varphi_{\eta\zeta} \,
  \frac{ \Delta(\eta, \zeta\myset{\xi}) }{ \myOmega{} \myOmega{\myset\xi} } 
  \; 
  \sum_{\xi\in\myset\xi}
  \frac{ \myOmega{\xi\bar\zeta} \myOmega{\eta\myset\xi/\xi} } 
       { \Delta(\xi, \eta\zeta\myset\xi/\xi) }.
\label{eq-ql-a}
\end{equation}
Using this equation for one element sets, $\myset\xi=\{\xi\}$, 
\begin{equation}
  \left( \myT{\xi} - 1 \right) f_{\eta\zeta}
  = 
  \varphi_{\eta\zeta} \,
  \frac{ \zeta - \eta }{ \xi - \zeta } \,
  \frac{ \myOmega{\xi\bar\zeta} \myOmega{\eta} } 
       { \myOmega{} \myOmega{\xi} }
\end{equation}
to substitute $\myOmega{\xi\bar\zeta}$ in \eref{eq-ql-a} it is easy to obtain 
\begin{equation}
  \left( \myT{\myset\xi} - 1 \right) f_{\eta\zeta} 
  = 
  \sum_{ \xi \in \myset\xi } 
  a_{\eta\xi\myset\xi} 
  \left( \myT\xi - 1 \right) f_{\eta\zeta}
\label{eq-gql-s}
\end{equation}
where 
\begin{equation}
  a_{\eta\xi\myset\xi} 
  = 
  \frac{ \Delta( \eta, \myset\xi/\xi ) }{ \Delta( \xi,  \myset\xi/\xi ) } \; 
  \frac{ \myOmega{\xi}  \myOmega{\eta\myset{\xi}/\xi} } 
       { \myOmega{\eta} \myOmega{\myset\xi} }. 
\end{equation}
Noting that $a_{\eta\xi\myset\xi}$ do not depend on $\zeta$, one can 
`vectorize' equation \eref{eq-gql-s} by introducing the vector 
$\vec{f}_{\eta}$ defined by 
\begin{equation}
  \vec{f}_{\eta} 
  = 
  \sum_{\zeta\in\myset\zeta} 
  f_{\eta\zeta} \, \vec{e}_{\zeta} 
\end{equation}
where $\{ \vec{e}_{\zeta} \}$ is an \textit{arbitrary} set of vectors from an 
\textit{arbitrary} space $\mathbb{V}$. Equation \eref{eq-gql-s} can now be 
rewritten as 
\begin{equation}
  \left( \myT{\myset\xi} - 1 \right) \vec{f}_{\eta} 
  = 
  \sum_{ \xi \in \myset\xi } 
  a_{\eta\xi\myset\xi} 
  \left( \myT\xi - 1 \right) \vec{f}_{\eta}. 
\label{eq-gql-v}
\end{equation}
It is clearly seen that the simplest case of \eref{eq-gql-v}, 
$\mysetsize\xi=2$, describes multidimensional quadrilateral lattices (QL)
in $\mathbb{V}$ \cite{BK95,DS97,BS08}.
Thus, we have demonstrated that the determinants $\Omega$ provide solutions 
for the generalized QL equation \eref{eq-gql-v}, that can be viewed as a 
compatible system of several classical (two-dimensional) QL equations.

Presenting $a_{\eta\xi\myset\xi}$ as 
\begin{equation}
  a_{\eta\xi\myset\xi} 
  = 
  \frac{ \myShifted1{b_{\eta\xi}}{\myset\xi/\xi} }{ b_{\eta\xi} } 
\end{equation} 
with 
\begin{equation}
  b_{\eta\xi} 
  = 
  - 
  \varphi_{\eta\xi} \, 
  \frac{ \myOmega\eta }{ \myOmega\xi } 
\end{equation}
one can conclude, after some simple manipulations with \eref{Fay-main},  
that $b_{\eta\xi}$ satisfies the equations 
\begin{equation}
  \left( \myT{\myset\xi} - 1 \right) b_{\eta\zeta} 
  = 
  \sum_{\xi\in\myset\xi} 
  \myShifted2{b_{\eta\xi}}{\zeta\myset\xi/\xi} b_{\xi\zeta} 
\end{equation}
that generalize the discrete Darboux equations, 
\begin{equation}
  \left( \myT{\xi} - 1 \right) b_{\eta\zeta} 
  = 
  \myShifted2{b_{\eta\xi}}{\zeta} b_{\xi\zeta} 
\end{equation}
(see \cite{BK95,DS97,BS08}).

The multidimensional QL and discrete Darboux equations were studied by 
various authors who obtained their Grammian- and Casorati-type determinant 
\cite{ LM98, WOGTS99}, 
the algebro-geometric 
\cite{AVK99,D07,D10}
and the soliton solutions \cite{NAH09, AN10}. 
The results presented in this section can be viewed as a simplification 
and generalization of the Cauchy matrix approach elaborated in \cite{NAH09, AN10}.

\subsection{Darboux equations. \label{sec-darboux}}

The next three examples are the differential systems, which are consequences of 
\eref{Fay-diff}.

The construction we use in this section is similar to \eref{def-ql-f},
\begin{equation}
  f_{\eta\zeta} = 
  \frac{ \varphi_{\eta\zeta} }{ \zeta-\eta }
  \frac{ \myT\eta\myT\zeta^{-1}\Omega }{ \Omega } 
\label{def-darboux-f}
\end{equation}
where $\eta$, $\zeta$ belong to an arbitrary set, 
\begin{equation}
  \eta, \zeta \in \myset\xi
\end{equation}
and functions $\varphi_{\eta\zeta}$ are solutions of  
\begin{equation}
  \partial_{\lambda} \ln\varphi_{\eta\zeta} 
  = 
  \frac{ 1 }{ \lambda - \eta } 
  - 
  \frac{ 1 }{ \lambda - \zeta }, 
  \qquad 
  \varphi_{\eta\eta} = 1. 
\end{equation}
It follows from \eref{Fay-diff} that functions \eref{def-darboux-f} satisfy 
\begin{equation}
  \partial_{\lambda} f_{\eta\zeta} 
  = 
  f_{\eta\lambda} f_{\lambda\zeta} 
\qquad
  (
  \lambda,\eta,\zeta \in \myset\xi, 
\quad
  \lambda \ne \eta \ne \zeta \ne \lambda 
  )
\end{equation}
which means that \eref{def-darboux-f} describe soliton solutions for the 
Darboux equations \cite{D,ZM85}.

\subsection{$(1+N)$-dimensional NLS-like system. \label{sec-app-C}}

The first two examples discussed above are `general' ones, in the sense that 
no restrictions are imposed on the matrices $\mymatrix{L}$ and $\mymatrix{R}$ 
determining, through \eref{cond-rank-one},  the matrices $\mymatrix{A}$ which are 
used to construct the determinants $\Omega$ . 
However, as can be seen from \cite{V13a,V05,PV11,V11,V12a,V13b}, 
to build solutions for the NLS-like system one has to use 
$\mymatrix{L}$ and $\mymatrix{R}$ that are not independent, which is a 
manifestation of the fact that dark solitons are one-parametric (contrary to 
the two-parametric bright ones). For example, in \cite{V05,V12a,V13b} 
$\mymatrix{L}=\mymatrix{R}^{-1}$. 
Here, we use generalization of this condition that leads to the generalized 
NLS-like equations.

Consider a fixed set of parameters 
\begin{equation}
  \myset\mu = \{ \mu_{1}, ..., \mu_{N} \} 
\end{equation}
and the restriction
\begin{equation}
  \myT{\myset\mu} = \myT\kappa 
\end{equation}
with fixed $\kappa$ which relates $\mymatrix{L}$ and $\mymatrix{R}$, 
\begin{equation}
  \mathcal{F}(\mymatrix{L}) 
  = 
  \mathcal{F}(\mymatrix{R}),
  \qquad
  \mathcal{F}(t) = 
  \frac{ \prod_{\mu\in\myset\mu} (t - \mu) }{ t - \kappa }.  
\label{level-f}
\end{equation}
The differential Fay identity \eref{Fay-diff} with $\myset\beta=\emptyset$ can be 
rewritten in terms of the Miwa shifts $\myE{}$ as 
\begin{equation}
  \biggl[ 
    D_{\kappa} 
    + \Lambda_{\kappa}(\myset\alpha) 
  \biggr] 
  \myOmega{\myset\alpha} \cdot \myOmega{} 
  =  
  \sum_{ \alpha \in \myset\alpha } 
  \gamma_{\alpha}(\myset\alpha) \; 
  \left( \myE\alpha \myOmega{} \right) 
  \left( \myE\alpha^{-1} \myOmega{\myset\alpha} \right) 
\end{equation}
where 
\begin{equation}
  \gamma_{\alpha}(\myset\alpha) =  
  \frac{ 1 }{ (\kappa-\alpha)^{2} } 
  \frac{ \Delta( \kappa, \myset\alpha ) }
       { \Delta( \alpha, \myset\alpha/\alpha ) }.
\end{equation}
Using this equation to calculate the derivative of 
$\myOmega{\myset\mu/\mu}/\myOmega{}$, 
\begin{equation}
  \biggl[ 
    \partial_{\kappa} 
    + \Lambda_{\kappa}(\myset\mu/\mu) 
  \biggr] 
  \frac{ \myOmega{\myset\mu/\mu} }{ \myOmega{} } 
  =  
  \sum_{ \alpha \in \myset\mu/\mu } 
  P_{\mu\alpha} \; 
  \myE\alpha^{-1} 
  \frac{ \myOmega{\myset\mu/\mu} }{ \myOmega{} } 
\end{equation}
where 
\begin{equation}
  P_{\mu\alpha} =  
  \gamma_{\alpha}(\myset\mu/\mu)
  \frac{ \left( \myE{\alpha}      \myOmega{} \right) 
         \left( \myE{\alpha}^{-1} \myOmega{} \right) }
       { \myOmega{}^{2} }, 
\end{equation}
and noting that 
\begin{equation}
  \myOmega{\myset\mu/\mu} = \myE{\mu}^{-1} \myOmega{} 
\end{equation}
one can present $P_{\mu\alpha}$ as a product of 
fractions $\myE\alpha^{\pm 1} \Omega / \Omega$ 
and constant factors. 
This leads to the closed system for the functions 
\begin{equation}
  q_{\mu} 
  = 
  \varphi_{\mu} \, 
  \frac{ \myE\mu^{-1} \Omega }{ \Omega }, 
\qquad
  r_{\mu} 
  = 
  \psi_{\mu} \, 
  \frac{ \myE\mu \Omega }{ \Omega } 	
\label{def-nls-qr} 
\end{equation}
that can be written as 
\begin{eqnarray} 
  \partial_{\kappa} q_{\mu} 
  & = & 
  \sum\limits_{ \nu \in \myset\mu/\mu} 
  q_{\nu} r_{\nu} \; 
  \myE\nu^{-1} q_{\mu}  
\label{syst-nls-q}
\\  
  -\partial_{\kappa} r_{\mu} 
  & = & 
  \sum\limits_{ \nu \in \myset\mu/\mu} 
  q_{\nu} r_{\nu} \; 
  \myE\nu r_{\mu} 
\label{syst-nls-r}
\end{eqnarray}
provided the background functions $\varphi_{\mu}$ and $\psi_{\mu}$ satisfy 
\begin{equation}
  \partial_{\kappa} \ln\varphi_{\mu} 
  = 
  - \partial_{\kappa} \ln\psi_{\mu} 
  = 
  \sum_{\nu\in\myset\mu/\mu} 
  \frac{ 1 }{ \kappa - \nu }, 
\end{equation}
\begin{equation}
  \frac{ \myE\nu \varphi_{\mu} }{ \varphi_{\mu} }  
  = 
  \frac{ \psi_{\mu} }{ \myE\nu \psi_{\mu} }  
  = 
  \frac{ \mu - \kappa }{ \mu - \nu } 
\end{equation}
and are related by 
\begin{equation}
  \varphi_{\mu} \psi_{\mu} = 
  \frac{ 1 }{ (\kappa-\mu)^{2} }
  \frac{ \Delta( \kappa, \myset\mu ) }
       { \Delta( \mu, \myset\mu/\mu ) } 
  = \mbox{constant} . 
\end{equation}

Now we pass to the vector notation 
by introducing, instead of $N$-set $\myset\mu$, a basis of $\mathbb{R}^{N}$
\begin{equation}
  \myset\mu = \{ \mu_{1}, ..., \mu_{N} \} 
  \; \to \; 
  \{ \vec{e}_{1}, ..., \vec{e}_{N} \} 
\end{equation}
and identifying the shift $\myE{\mu_{i}}$ with translation by $\vec{e}_{i}$. 
In new terms, system \eref{syst-nls-q}, \eref{syst-nls-r} can be written as 
\begin{eqnarray}
  \phantom{-}\dot{q}_{i}(\vec{n}) 
  & = & 
  \sum\limits_{ a \ne i } 
  q_{a}(\vec{n}) r_{a}(\vec{n}) \; 
  q_{i}(\vec{n}-\vec{e}_{a}) 
\label{eqs-nls-q}
\\
  - \dot{r}_{i}(\vec{n}) 
  & = & 
  \sum\limits_{a \ne i} 
  q_{a}(\vec{n}) r_{a}(\vec{n}) \; 
  r_{i}(\vec{n}+\vec{e}_{a})
\label{eqs-nls-r}
\end{eqnarray}
where we write $q_{i}$ instead of $q_{\mu_{i}}$ 
and use dot to indicate the derivative, 
$\dot{q} = \partial_{\kappa} q$, \textit{etc}.

To summarize, we have demonstrated that functions defined by \eref{def-nls-qr} 
are solutions of the $(1+N)$-dimensional NLS-like system \eref{eqs-nls-q} and 
\eref{eqs-nls-r}.
To present the $N_{s}$-soliton solutions for this model one has to rewrite 
formulae of this section, together with the `evolution' equation for the 
matrix $\mymatrix{A}$ following from \eref{partial-def},
\begin{equation}
  \partial_{\kappa} \mymatrix{A} 
  = 
  \mymatrix{A} \cdot
  (\mymatrix{L}-\mymatrix{R})
  (\mymatrix{L}-\kappa)^{-1}  
  (\mymatrix{R}-\kappa)^{-1},  
\end{equation}
in the $t$-$\vec{n}$ notation. 
Omitting the details (such as using \eref{cond-rank-one} to write the 
elements of $\mymatrix{A}$, introducing $\exp(\myu_{i})$ instead of 
$\varphi_{\mu_{i}}$ etc), we present here the `final' form of 
$q_{i}$ and $r_{i}$:
\begin{eqnarray}
  q_{i}(t,\vec{n}) 
  & = &
  q_{i}^{\scriptscriptstyle(0)}
  e^{ \myu_{i}(t,\vec{n}) }
  \frac{\det\left| 
    \mymatrix{1} + \mymatrix{A}\left(t,\vec{n}-\vec{e}_{i}\right)
  \right|}
  {\det\left| 
    \mymatrix{1} + \mymatrix{A}\left(t,\vec{n}\right)
  \right|}
\\
  r_{i}(t,\vec{n}) 
  & = &
  r_{i}^{\scriptscriptstyle(0)}
  e^{ -\myu_{i}(t,\vec{n}) }
  \frac{\det\left| 
    \mymatrix{1} + \mymatrix{A}\left(t,\vec{n}+\vec{e}_{i}\right)
  \right|}
  {\det\left| 
    \mymatrix{1} + \mymatrix{A}\left(t,\vec{n}\right)
  \right|}.
\end{eqnarray}
Here, 
$\mymatrix{A}$ is the $N_{s} \times N_{s}$ matrix, 
$\mymatrix{A} = \left( A_{\alpha\beta} \right)_{\alpha,\beta=1..N_{s}}$, 
with the elements 
\begin{equation}
  A_{\alpha\beta}(t,\vec{n}) 
  = 
  \frac{ 1 }{ L_{\alpha} - R_{\beta} } 
  \exp\left\{ \myw_{\beta}(t,\vec{n}) \right\},
  \qquad 
  \alpha,\beta=1,...,N_{s} 
\end{equation}
where the pairs $L_{\alpha}$ and $R_{\alpha}$ belong to the same level set 
of the function $\mathcal{F}$ from \eref{level-f}: 
$L_{\alpha}$ and $R_{\alpha}$ are distinct solutions of 
$\mathcal{F}(t)=C_{\alpha}$ for arbitrary $C_{\alpha}$.
The `phases' $\myu_{i}$ and $\myw_{\beta}$ are given by 
\begin{eqnarray}
  \myu_{i}(t,\vec{n}) 
  & = & 
  \myu_{i0}t + \left( \vec{\myu_{i}},\vec{n} \right) + \mbox{constant} 
\\
  \myw_{\beta}(t,\vec{n}) 
  & = & 
  \myw_{\beta 0}t + \left( \vec{\myw_{\beta}},\vec{n} \right) + \mbox{constant} 
\end{eqnarray} 
with the `frequencies'
\begin{eqnarray}
  \myu_{i0} 
  & = & 
  \frac{ 1 }{ \mu_{i} - \kappa } 
  - \sum_{j=1}^{N}\frac{ 1 }{ \mu_{j} - \kappa }
\\
  \myw_{\beta 0} 
  & = & 
  \frac{ 1 }{ R_{\beta} - \kappa } 
  - 
  \frac{ 1 }{ L_{\beta} - \kappa } 
\end{eqnarray}
and the `space' vectors 
\begin{eqnarray}
  \vec{\myu}_{i} 
  & = & 
  \sum_{a \ne i}
  \ln \frac{ \mu_{i} - \kappa }{ \mu_{i} - \mu_{a} } \; 
  \vec{e}_{a} 
\\
  \vec{\myw}_{\beta} 
  & = & 
  \sum_{i=1}^{N}
  \ln \frac{ \mu_{i} - L_{\beta} }{ \mu_{i} - R_{\beta} } 
      \frac{ \kappa - R_{\beta} }{ \kappa - L_{\beta} } \; 
  \vec{e}_{i} 
\end{eqnarray} 
while the constants 
$q_{i}^{\scriptscriptstyle(0)}$, $r_{i}^{\scriptscriptstyle(0)}$ 
are related by 
\begin{equation}
  q_{i}^{\scriptscriptstyle(0)} r_{i}^{\scriptscriptstyle(0)} 
  = 
  \frac{ \exp\left\{ 
           \left( \vec{\myu}_{tot} - \vec{\myu}_{i}, \vec{e}_{i} \right) 
         \right\} }
       { \kappa - \mu_{i} },
  \qquad
  \vec{\myu}_{tot} = \sum_{j=1}^{N} \vec{\myu}_{j}. 
\end{equation}
In the above formulae, we use $(\vec{a},\vec{b})$ to denote the standard 
scalar product of vectors $\vec{a}$ and $\vec{b}$ 
and presume that basis $\{ \vec{e}_{i} \} $ is orthogonal.

To conclude this section we would like to comment the dark-soliton character 
of the presented solution. It is easy to see from the structure of the matrix 
$\mymatrix{A}$ that for the vector $\vec{n}_{0}$ such that 
$\exp\{ (\vec{\myw}_{\beta}, \vec{n}_{0} )\}$ is, say, positive for all $\beta$ 
the elements of $\mymatrix{A}(t,k\vec{n}_{0})$ tend to $0$ 
(when $k \to -\infty$) or to the infinity (when $k \to +\infty$). Thus, 
$\det\left| \mymatrix{1}+\mymatrix{A}(t,k\vec{n}_{0}) \right|$ tends 
to $1$ or is given by $\det\left| \mymatrix{A}(t,k\vec{n}_{0}) \right| + O(1)$, which 
implies that 
\begin{equation}
  q_{i}(t, k\vec{n}_{0}) 
  \sim 
  q_{i}^{\scriptscriptstyle(0)}
  e^{ \myu_{i}(t,\vec{n}) }
  \left\{
    \begin{array}{lcl}
    1 & & k \to - \infty \\
    e^{-\left( \vec{w}_{tot}, \vec{e}_{i} \right)}
    & & k \to + \infty
    \end{array}
  \right.
\end{equation}
where $\vec{w}_{tot}=\sum_{j=1}^{N}\vec{w}_{j}$.
Interpreting $e^{ \myu_{i}}$ as a background (it can be made usual plane wave 
by taking pure imaginary $\myu_{i}$) one can see the typical `dark soliton' 
picture: a background solution in the asymptotic regions, nontrivial behaviour 
in the region of finite $\vec{n}$'s and a constant `phaseshift',
$-\left( \vec{w}_{tot}, \vec{e}_{i} \right)$.

\subsection{(1+2)-dimensional lattice with saturable nonlinearity. 
\label{sec-app-D}} 

In this example we slightly modify calculations of the previous section and 
discuss a model with saturating nonlinearity.

Consider three shifts $\myT\xi$, $\xi \in \{\mu, \nu, \kappa \}$ related by 

\begin{equation}
  \myT\mu \myT\nu = \myT\kappa. 
\label{mrt-restr}
\end{equation}
The differential Fay identity \eref{Fay-diff} with 
$\myset\alpha = \{\mu,\nu\}$ and $\myset\beta = \emptyset$ under restriction 
\eref{mrt-restr} reads 
\begin{equation}
  \left( \partial_{\kappa} + \Lambda_{\kappa} \right) 
  \frac{ \hat\Omega }{\Omega} 
  = 
  \frac{ P_{\mu} }{ \kappa - \nu } \myT\mu \frac{\hat\Omega}{\Omega} 
  + 
  \frac{ P_{\nu} }{ \kappa - \mu } \myT\nu \frac{\hat\Omega}{\Omega} 
\label{mrt-aa}
\end{equation} 
or, after applying $\myT\kappa^{-1}$, 
\begin{equation}
  \left( -\partial_{\kappa} + \Lambda_{\kappa} \right) 
  \frac{ \check\Omega }{\Omega} 
  = 
  \frac{ P_{\mu} }{ \kappa - \nu } \myT\mu^{-1} \frac{\check\Omega}{\Omega} 
  + 
  \frac{ P_{\nu} }{ \kappa - \mu } \myT\nu^{-1} \frac{\check\Omega}{\Omega} 
\label{mrt-ab}
\end{equation} 
with $\Lambda_{\kappa}=\Lambda_{\kappa}(\{\mu,\nu\})$,
\begin{equation}
  \Lambda_{\kappa} = \frac{1}{\kappa-\mu} + \frac{1}{\kappa-\nu}. 
\end{equation}
Here we again use notation \eref{def-hat}, 
$\hat\Omega = \myT\kappa \Omega$ and $\check\Omega = \myT\kappa^{-1}\Omega$, 
while $P_{\mu,\nu}$ are given by 
\begin{equation}
  P_{\mu} 
  = 
  \frac{\mu-\kappa}{\mu-\nu} \, 
  \frac{ \left( \myT\mu \Omega \right)\left( \myT\mu^{-1} \Omega \right) }
       {\Omega^{2}}, 
  \qquad
  P_{\nu} 
  = 
  \frac{\nu-\kappa}{\nu-\mu} \, 
  \frac{ \left( \myT\nu \Omega \right)\left( \myT\nu^{-1} \Omega \right) }
       {\Omega^{2}}. 
\end{equation}
In terms of new functions
\begin{equation}
  q = \varphi \, \frac{ \hat\Omega }{ \Omega }, 
  \qquad
  r = \frac{1}{\varphi} \, \frac{ \check\Omega }{ \Omega } 
\label{mrt-qr}
\end{equation} 
where $\varphi$ is determined by 
\begin{equation}
  \partial_{\kappa} \ln \varphi = \Lambda_{\kappa},
  \qquad
  \frac{ \myT\mu \varphi }{ \varphi }  
  = 
  \sqrt\frac{\mu-\kappa}{\nu-\kappa}, 
  \qquad
  \frac{ \myT\nu \varphi }{ \varphi }  
  = 
  \sqrt\frac{\nu-\kappa}{\mu-\kappa}  
\end{equation}
equations \eref{mrt-aa} and \eref{mrt-ab} can be written as 
\begin{eqnarray}
  - \rho\,\partial_{\kappa} q 
  & = & 
  P_{\mu} \, \myT\mu \, q + P_{\nu} \, \myT\nu \, q 
\\[2mm]
  \rho\,\partial_{\kappa} r 
  & = & 
  P_{\mu} \, \myT\mu^{-1} r + P_{\nu} \, \myT\nu^{-1} r  
\end{eqnarray}
where
\begin{equation}
  \rho = 
  \sqrt{(\mu-\kappa)(\nu-\kappa)}. 
\end{equation}
To close this system one has to relate $P_{\mu}$ and $P_{\nu}$  with $q$ and 
$r$. This can be done using the observation  
\begin{equation}
  \myT\nu \Omega = \myT\mu^{-1} \hat\Omega, 
  \qquad
  \myT\nu^{-1} \Omega = \myT\mu \check\Omega 
\end{equation}
which leads to 
\begin{equation}
  P_{\nu} 
  = 
  - 
  P_{\mu} \, \left( \myT\mu^{-1} q \right) \left( \myT\mu r \right). 
\label{mrt-c}
\end{equation}
This, together with 
\begin{equation}
  P_{\mu} + P_{\nu} = 1, 
\label{mrt-b}
\end{equation}
that follows from the main Fay identity \eref{Fay-main} for  
$\myset\xi = \{\kappa\}$ and $\myset\eta = \{\mu,\nu\}$ 
after applying the shift $\myT\kappa^{-1}$, 
implies  
\begin{equation}
  P_{\mu} = 
  \frac{ 1 }{ 1 - \left(\myT\mu^{-1} q \right)\left(\myT\mu r \right) }, 
  \qquad
  P_{\nu} = 
  \frac{ 1 }{ 1 - \left(\myT\nu^{-1} q \right)\left(\myT\nu r \right) }. 
\end{equation}

Finally, introducing the `time' by 
$ d/dt = \rho\partial_{\kappa}$ 
and using the vector notation (as in the previous section) one arrives at the 
following result: functions $q$ and $r$ solve the hyperbolic 
(or NLS-like) system
\begin{eqnarray}
  - \dot{q}(\vec{n}) 
  & = & 
  \sum_{i=1,2}
  \frac{ q\left(\vec{n}+\vec{e}_{i} \right) }
       { 1 - q\left( \vec{n} - \vec{e}_{i} \right) 
             r\left( \vec{n} + \vec{e}_{i} \right) } 
\label{eq-dot-q}
\\ 
  \phantom{-}\dot{r}(\vec{n}) 
  & = & 
  \sum_{i=1,2}
  \frac{ r\left( \vec{n} - \vec{e}_{i} \right) }
       { 1 - q\left( \vec{n} - \vec{e}_{i} \right) 
             r\left( \vec{n} + \vec{e}_{i} \right) } 
\label{eq-dot-r}
\end{eqnarray} 
where dot stands for $d/dt$.

To conclude, we write the `final' formulae for the $N_{s}$-soliton 
solutions. We restrict ourselves with the case $\kappa=0$ 
(other choice, $\kappa \ne 0$, leads to rescaling of the time and redefinition 
of constant parameters). Condition \eref{mrt-restr} leads to 
\begin{equation}
  \nu=\mu^{-1},
  \qquad
  \mymatrix{R}=\mymatrix{L}^{-1},
  \qquad
  \rho=1.
\end{equation}
Noting that $\mymatrix{H}_{\kappa}=\mymatrix{L}^{2}$ one can write $q$ and $r$ 
in the following form:
\begin{eqnarray}
  q(t,\vec{n}) 
  & = &
  e^{ \myu(t,\vec{n}) }
  \frac{\det\left| 
    \mymatrix{1} + \mymatrix{A}\left(t,\vec{n}\right)\mymatrix{L}^{2}
  \right|}
  {\det\left| 
    \mymatrix{1} + \mymatrix{A}\left(t,\vec{n}\right)
  \right|}
\\
  r(t,\vec{n}) 
  & = &
  e^{ -\myu(t,\vec{n}) }
  \frac{\det\left| 
    \mymatrix{1} + \mymatrix{A}\left(t,\vec{n}\right)\mymatrix{L}^{-2}
  \right|}
  {\det\left| 
    \mymatrix{1} + \mymatrix{A}\left(t,\vec{n}\right)
  \right|}.
\end{eqnarray}
The elements of the $N_{s} \times N_{s}$ matrix $\mymatrix{A}$ are given by 
\begin{equation}
  A_{\alpha\beta}(t,\vec{n}) 
  = 
  \frac{ 1 }{ 1 - L_{\alpha}L_{\beta} } 
  \exp\left\{ \myw_{\beta}(t,\vec{n}) \right\},
  \qquad 
  \alpha,\beta=1,...,N_{s} 
\end{equation}
where 
\begin{equation}
  \myw_{\beta}(t,\vec{n}) 
  = 
  \left( L_{\beta} - L_{\beta}^{-1} \right) t 
  + \left( \vec{\myw_{\beta}},\vec{n} \right) + \mbox{constant} 
\end{equation} 
with 
\begin{equation}
  \vec{\myw}_{\beta} 
  = 
  \ln L_{\beta} \; \vec{\varepsilon} 
  + \ln \frac{ L_{\beta} - \mu }{ 1 - \mu L_{\beta} } \; 
  \sigma_{3}\vec{\varepsilon} 
\end{equation} 
and 
\begin{equation}
  \myu(t,\vec{n}) 
  = 
  \left( \vec{\mu}, \vec{n} - \vec{\varepsilon}t \right) 
  + 
  \mbox{constant}. 
\end{equation} 
Here, $\vec{\mu}$, $\vec{\varepsilon}$ are the constant vectors
\begin{equation}
  \vec{\mu} = 
  \left( \begin{array}{c} \mu \\ 1/\mu \end{array} \right), 
\qquad
  \vec{\varepsilon} = 
  \left( \begin{array}{c} 1 \\ 1 \end{array} \right)
\end{equation} 
and $\sigma_{3}$ is the Pauli matrix $\mbox{diag}\left(1,-1\right)$.

\section{Conclusion.}

The main result of this work is the set of \textit{bilinear} identities 
that can be viewed as soliton analogues of the Fay identities 
and which, in our opinion, are useful from the practical viewpoint. 
First, they simplify the procedure of constructing soliton solutions. It has 
been demonstrated in the previous works of the author as well as in examples 
provided above that starting from \eref{Fay-main}, \eref{Fay-diff} or the 
formulae from section \ref{sec-miwa}, one can obtain by simple algebra
explicit solutions for a wide range of equations such as, \textit{e.g}, 
equations from the NLS hierarchy \cite{V12a}, from the derivative NLS 
hierarchy  \cite{V13b}, equations that describe multidimensional QL, discrete 
and differential Darboux equations (see examples from sections \ref{sec-mql} 
and \ref{sec-darboux}) \textit{etc}.

Another application of the obtained Fay identities is for a search for new 
solvable models, especially in multidimensions. Two examples of apparently new 
equations can be found in sections \ref{sec-app-C} and \ref{sec-app-D}. Of 
course, we cannot, at present, say anything about integrability of 
\eref{eqs-nls-q}, \eref{eqs-nls-r}
or \eref{eq-dot-q} and \eref{eq-dot-r}. However, the fact that these systems 
admit the $N$-soliton solutions is an indication that they are integrable 
(or at least admit integrable reductions, as in the case studied in 
\cite{PV10}). 
Surely, examples from sections \ref{sec-app-C} and \ref{sec-app-D} 
do not exhaust all interesting (both from the viewpoint of integrability and 
of possible physical applications) models that can be solved by means of the 
identities discussed in this paper and we hope to find more in future. 

Finally, we would like to mentioned the most straightforward continuation of 
this work: to derive the Fay identities for the \textit{bright} soliton case.
It is interesting that usually the dark solitons are considered as more 
`difficult' objects than the bright ones. This is probably because of the 
fact that to derive them in the framework of the inverse scattering transform 
one has to study the spectral problems with nonvanishing at the infifnity 
potentials, which indeed complicates the involved mathematics. However, in 
the framework of the direct approach, the situation can be inverted. 
For example, the structure of the 
$\sigma$-$\tau$-$\rho$ triplet for the complex models turns out to be more 
complicated in the bright soliton case than the 
$\hat\Omega$-$\Omega$-$\check\Omega$ construction that appears in section 
\ref{sec-miwa}. This and other questions related to the bright soliton 
Fay identities will be discussed in the forthcoming paper.

\appendix
\section{Proof of proposition \ref{prop-PN}. \label{proof-PN}}

The aim of this section is to prove equations \eref{eq-omega-p} and 
\eref{eq-omega-n}, which are straightforward consequences of 
\eref{cond-rank-one}. Considering the first of them, one can write 
\begin{eqnarray}
  \myT\zeta \Omega(\mymatrix{A}) 
  & = & 
  \det\left|
    \mymatrix{1} + 
    \mymatrix{A} 
    \left(\mymatrix{L} - \zeta\right)
    \left(\mymatrix{R} - \zeta\right)^{-1} 
  \right|
\nonumber 
\\[2mm]
  & = & 
  \det\left|
    \mymatrix{1} + 
    \left(\mymatrix{L} - \zeta\right)
    \mymatrix{A} 
    \left(\mymatrix{R} - \zeta\right)^{-1} 
  \right|
\nonumber 
\\[2mm]
  & = & 
  \det\left|
    \mymatrix{1} + 
    \left( 
      \mymatrix{A} \left(\mymatrix{R} - \zeta\right) 
      + | \,\ell\, \rangle\langle a | 
    \right)
    \left(\mymatrix{R} - \zeta\right)^{-1} 
  \right|
\nonumber 
\\[2mm]
  & = & 
  \det\left|
    \mymatrix{1} + 
    \mymatrix{A} + 
    | \,\ell\, \rangle\langle a | \left(\mymatrix{R} - \zeta\right)^{-1} 
  \right|
\nonumber 
\\[2mm]
  & = & 
  \Omega(\mymatrix{A}) 
  \det\left|
    \mymatrix{1} + 
    G(\mymatrix{A}) 
    | \,\ell\, \rangle\langle a | \left(\mymatrix{R} - \zeta\right)^{-1} 
  \right|.
\end{eqnarray}
The last expression, together with 
\begin{equation}
  \det\left|
    \mymatrix{1} +  | \,u\, \rangle\langle v | 
  \strut\right|
  = 
  1 + \langle v | \,u\, \rangle
\end{equation}
leads to \eref{eq-omega-p}.

In a similar way, 
\begin{eqnarray}
  \myT\zeta^{-1} \Omega(\mymatrix{A}) 
  & = & 
  \det\left|
    \mymatrix{1} + 
    \mymatrix{A} 
    \left(\mymatrix{R} - \zeta\right)
    \left(\mymatrix{L} - \zeta\right)^{-1} 
  \right|
\nonumber 
\\[2mm]
  & = & 
  \det\left|
    \mymatrix{1} + 
    \left(\mymatrix{L} - \zeta\right)^{-1} 
    \mymatrix{A} 
    \left(\mymatrix{R} - \zeta\right)
  \right|
\nonumber 
\\[2mm]
  & = & 
  \det\left|
    \mymatrix{1} + 
    \left(\mymatrix{L} - \zeta\right)^{-1} 
    \left( 
      \left(\mymatrix{L} - \zeta\right) \mymatrix{A} 
      - | \,\ell\, \rangle\langle a | 
    \right)
  \right|
\nonumber 
\\[2mm]
  & = & 
  \det\left|
    \mymatrix{1} + 
    \mymatrix{A} - 
    \left(\mymatrix{L} - \zeta\right)^{-1} 
    | \,\ell\, \rangle\langle a | 
  \right|
\nonumber 
\\[2mm]
  & = & 
  \Omega(\mymatrix{A}) 
  \det\left|
    \mymatrix{1} - 
    G(\mymatrix{A}) 
    \left(\mymatrix{L} - \zeta\right)^{-1} 
    | \,\ell\, \rangle\langle a | 
  \right|
\end{eqnarray}
which implies \eref{eq-omega-n}.

\section{Proof of proposition \ref{prop-AB}. \label{proof-AB}}

To prove \eref{dets-XY}, we use \eref{eq-omega-p} and \eref{eq-omega-n} 
in the following form 
\begin{eqnarray}
  \myT\zeta \Omega(\mymatrix{X}) 
  & = & 
  \Omega(\mymatrix{X}) 
  \left[
  1 + 
  \langle x | (\mymatrix{R} - \zeta)^{-1} \mymatrix{G}(\mymatrix{X}) | \,\ell\, \rangle
  \right]
\\[2mm]
  \myT\zeta^{-1} \Omega(\mymatrix{Y}) 
  & = & 
  \Omega(\mymatrix{Y}) 
  \left[
  1 - 
  \langle y | \mymatrix{G}(\mymatrix{Y}) (\mymatrix{L} - \zeta)^{-1} | \,\ell\, \rangle
  \right]
\end{eqnarray}
where 
\begin{equation}
  \langle x |  = \langle a |  \mymatrix{H}_{\myset\xi}, 
  \qquad
  \langle y |  = \langle a |  \mymatrix{H}_{\myset\eta} 
\end{equation}
and present 
\begin{eqnarray}
  \mathcal{F}_{\myset{\xi}\myset{\eta}}(\zeta) 
  & = & 
  \frac{ 
    \left(\myT\zeta \Omega(\mymatrix{X}) \right)
    \left(\myT\zeta^{-1} \Omega(\mymatrix{Y}) \right) }
  { \Omega(\mymatrix{X}) \Omega(\mymatrix{Y}) }
\nonumber 
\\
  & = & 
  \left[
  1 - 
  \langle y | \mymatrix{G}(\mymatrix{Y}) (\mymatrix{L} - \zeta)^{-1} | \,\ell\, \rangle
  \right]
  \left[
  1 + 
  \langle x | (\mymatrix{R} - \zeta)^{-1} \mymatrix{G}(\mymatrix{X}) | \,\ell\, \rangle
  \right]
\end{eqnarray}
as 
\begin{equation}
  \mathcal{F}_{\myset{\xi}\myset{\eta}}(\zeta) 
  = 
  1 
  + 
  \langle y | \mymatrix{G}(\mymatrix{Y}) 
  D_{\myset{\xi}\myset{\eta}}(\zeta) 
  \mymatrix{G}(\mymatrix{X}) 
  | \,\ell\, \rangle.
\label{eq-FD}
\end{equation}
To do this we use, when necessary, 
\begin{equation}
  \langle x | 
  = 
  \langle y | \mymatrix{H}_{\myset\xi} \mymatrix{H}_{\myset\eta}^{-1} 
  = 
  \langle y | 
  \mymatrix{G}(\mymatrix{Y}) 
  \left( \mymatrix{1} + \mymatrix{Y} \right) 
  \mymatrix{H}_{\myset\xi} \mymatrix{H}_{\myset\eta}^{-1} 
\end{equation}
and 
\begin{equation}
  | \,\ell\, \rangle 
  = 
  \left( \mymatrix{1} + \mymatrix{X} \right) 
  \mymatrix{G}(\mymatrix{X}) 
  | \,\ell\, \rangle.
\end{equation}
Thus,
\begin{eqnarray}
  D_{\myset{\xi}\myset{\eta}}(\zeta) 
  & = & 
  \left( \mymatrix{1} + \mymatrix{Y} \right) 
  \mymatrix{H}_{\myset\xi} \mymatrix{H}_{\myset\eta}^{-1} 
  \left( \mymatrix{R} - \zeta \right)^{-1} 
  - 
  \left( \mymatrix{L} - \zeta \right)^{-1} 
  \left( \mymatrix{1} + \mymatrix{X} \right) 
\nonumber 
\\&& 
  - 
  \left( \mymatrix{L} - \zeta \right)^{-1} 
  | \,\ell\, \rangle\langle x | 
  \left( \mymatrix{R} - \zeta \right)^{-1}. 
\end{eqnarray}
Replacing 
  $\mymatrix{Y} \mymatrix{H}_{\myset\xi} \mymatrix{H}_{\myset\eta}^{-1}$ 
with 
  $\mymatrix{X}$ 
one arrives at 
\begin{eqnarray}
  D_{\myset{\xi}\myset{\eta}}(\zeta) 
  & = & 
  \mymatrix{H}_{\myset\xi} \mymatrix{H}_{\myset\eta}^{-1} 
  \left( \mymatrix{R} - \zeta \right)^{-1} 
  - 
  \left( \mymatrix{L} - \zeta \right)^{-1} 
\nonumber 
\\&& 
  + 
  \left( \mymatrix{L} - \zeta \right)^{-1} 
  \left[
    \left( \mymatrix{L} - \zeta \right) \mymatrix{X} 
    - 
    \mymatrix{X} \left( \mymatrix{R} - \zeta \right) 
    - 
    | \,\ell\, \rangle\langle x | 
  \right]
  \left( \mymatrix{R} - \zeta \right)^{-1} 
\end{eqnarray}
Noting that the expression inside the square brackets vanishes by virtue of 
\eref{cond-rank-one} one can rewrite \eref{eq-FD} as 
\begin{equation}
  \mathcal{F}_{\myset{\xi}\myset{\eta}}(\zeta) 
  = 
  1 
  + 
  \langle y | \mymatrix{G}(\mymatrix{Y}) 
  \mathcal{D}_{\myset{\xi}\myset{\eta}}(\zeta) 
  \mymatrix{G}(\mymatrix{X}) 
  | \,\ell\, \rangle 
\end{equation}
which ends the proof of \eref{dets-XY}.


\end{document}